\documentclass[a4paper,11pt]{article}

\usepackage{pos}
\usepackage{lipsum} 
\usepackage[separate-uncertainty=true]{siunitx}

\title{Approximating new ice models with B-splines for improved IceCube event reconstruction: application to cascades and tracks}

\ShortTitle{Improved IceCube event reconstruction}

\author{The IceCube Collaboration \\{\normalsize \normalfont(a complete list of authors can be found at the end of the proceedings)}\\}

\emailAdd{tyuan@icecube.wisc.edu}

\abstract{

Event signatures in IceCube are complex, modulated by both particle physics and properties of the ice and detector. Event reconstruction thus requires accurate modeling of ice properties and detector effects to fit for physics parameters, such as energy and direction. Here, we highlight how improvements in calibration can translate into substantially improving the angular resolution of electromagnetic showers. Since showers are also used to model stochastic energy losses of tracks, we further show how improved ice modeling, along with other track-specific optimizations, leads to more meaningful directional likelihood spaces for high-energy muons. The median angular resolution for showers is improved by a factor of two over an older B-spline model, and accurate directional contours for tracks can be obtained with Wilks’ theorem.

\vspace{4mm}
{\bfseries Corresponding authors:}
Tianlu Yuan$^{1*}$\\
{$^{1}$ \itshape Dept. of Physics and Wisconsin IceCube Particle Astrophysics Center, University of Wisconsin{\textendash}Madison}\\[4mm]
$^*$ Presenter

\ConferenceLogo{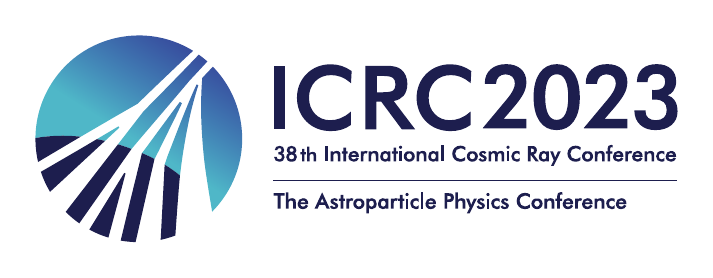}

\FullConference{The 38th International Cosmic Ray Conference (ICRC2023)\\ 26 July -- 3 August, 2023\\ Nagoya, Japan}
}

\newcommand{\zff}{z_{\text{eff}}}
\newcommand{\zen}{\Theta_{\text{zen}}}
\newcommand{\azi}{\Phi_{\text{azi}}}
\newcommand{\edep}{E_{\text{dep}}}

\begin{document}

\maketitle

\section{Introduction}\label{sec:intro}

The IceCube Neutrino Observatory is a cubic-kilometer array of digital optical modules (DOM) each instrumented with a single downward facing photomultiplier tube (PMT)~\cite{Aartsen:2016nxy}. Charged secondary particles from in-ice neutrino interactions and atmospheric muons that traverse the overburden above the detector both produce Cherenkov radiation detectable by IceCube PMTs. Though the instrumentation is sparse, the detected photoelectron signatures are imprinted with information on the physics parameters of interest: $\edep$, the visible energy, and $(\zen, \azi)$, the arrival direction.

Accurate event reconstruction requires accurate modeling of photon arrival time distributions for a given physics hypothesis; the signature of each event is highly dependent on the underlying neutrino interaction channel and resultant charged-lepton or hadron induced particle showers. In IceCube there are, broadly speaking, three general event categories: tracks, cascades and double cascades arising predominantly from muons, electromagnetic (EM) or hadronic showers and tau neutrino charged current interactions, respectively. High-energy muons deposit energy losses stochastically along their travel direction and high-energy taus decay leptonically or semi-leptonically. Ultimately the dominant contribution of detected Cherenkov photons are due to particle showers. Thus, a reasonable atomic unit for all reconstruction is the cascade model, with high-energy tracks and double cascades generalizing to employ models that include multiple, constrained cascades~\cite{Aartsen:2013vja,hallen2013measurement}.

In this proceeding, I discuss some recent improvements in the modeling of shower Cherenkov photon yields with B-splines~\cite{Whitehorn:2013nh}. Section \textsection\ref{sec:model} describes how updates to the layer undulations in the South Pole ice can be approximated. Section \textsection\ref{sec:benchmarks} discusses applications and benchmarks using simulated cascades. Section \textsection\ref{sec:realtime} discusses improvements for realtime alerts. Section \textsection\ref{sec:outro} touches on areas of potential future improvement and concludes.

\section{Approximating the new ice tilt}\label{sec:model}
Photon arrival time distributions depend on ice characteristics. Due to inhomogeneities in glacial ice sheet~\cite{IceCube:2013llx} an analytic description is challenging. Instead, IceCube has modeled photon arrival time distributions with B-splines~\cite{Aartsen:2013vja} and, more recently, with neural networks~\cite{IceCube:2021umt}. Recent advances in ice modeling include updates to the description of ice anisotropy~\cite{tc-2022-174} and layer undulations or ice tilt~\cite{Chirkin:2023icrc}. A approximate description of the former can be achieved using the concept of an effective distance correction~\cite{Usner:2018cel,YUAN2023168440}. An approximate description of the latter can be achieved using an effective depth correction~\cite{YUAN2023168440}.
\begin{figure*}[hbt]
\centering
\includegraphics[width=0.49\textwidth]{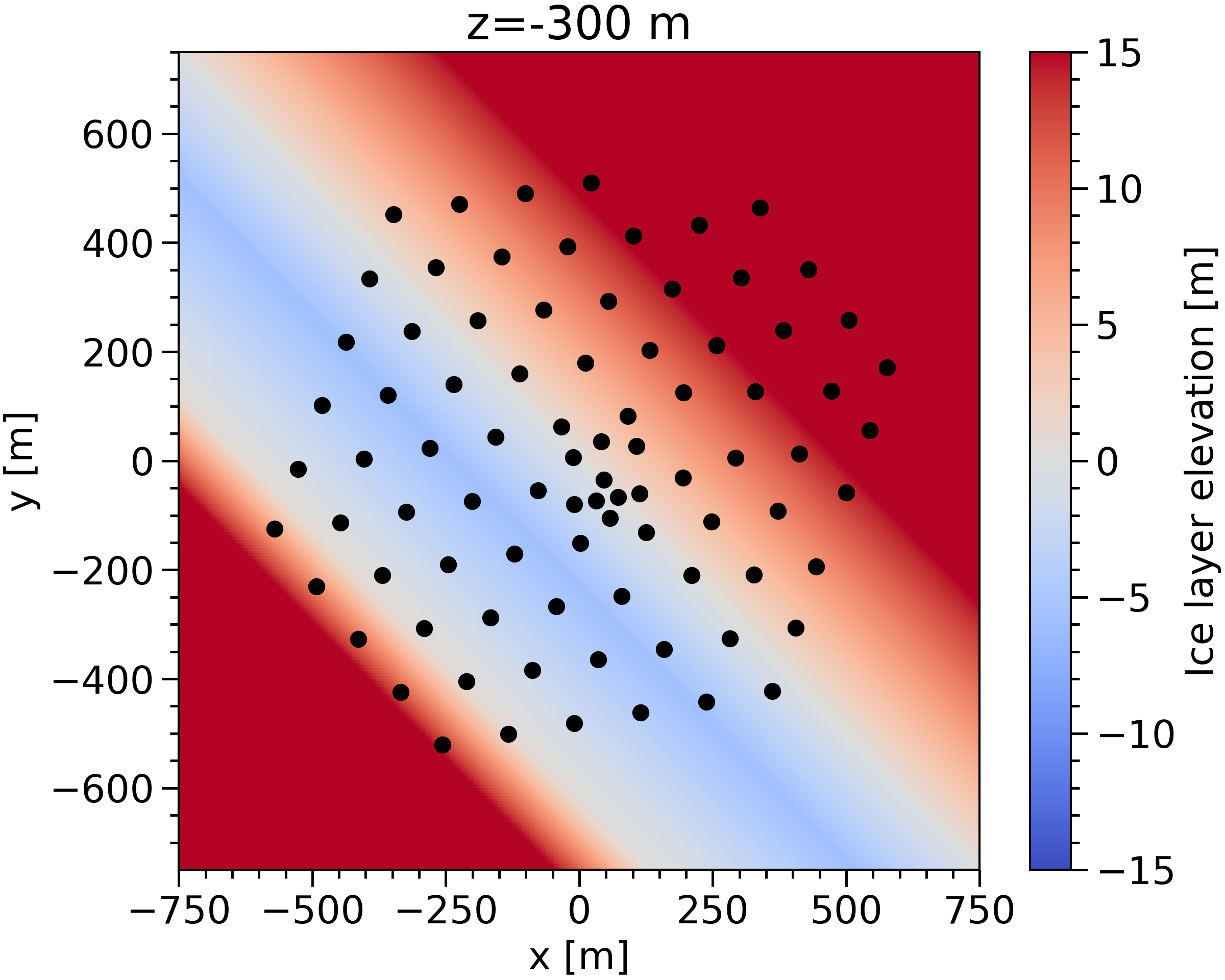}
\includegraphics[width=0.49\textwidth]{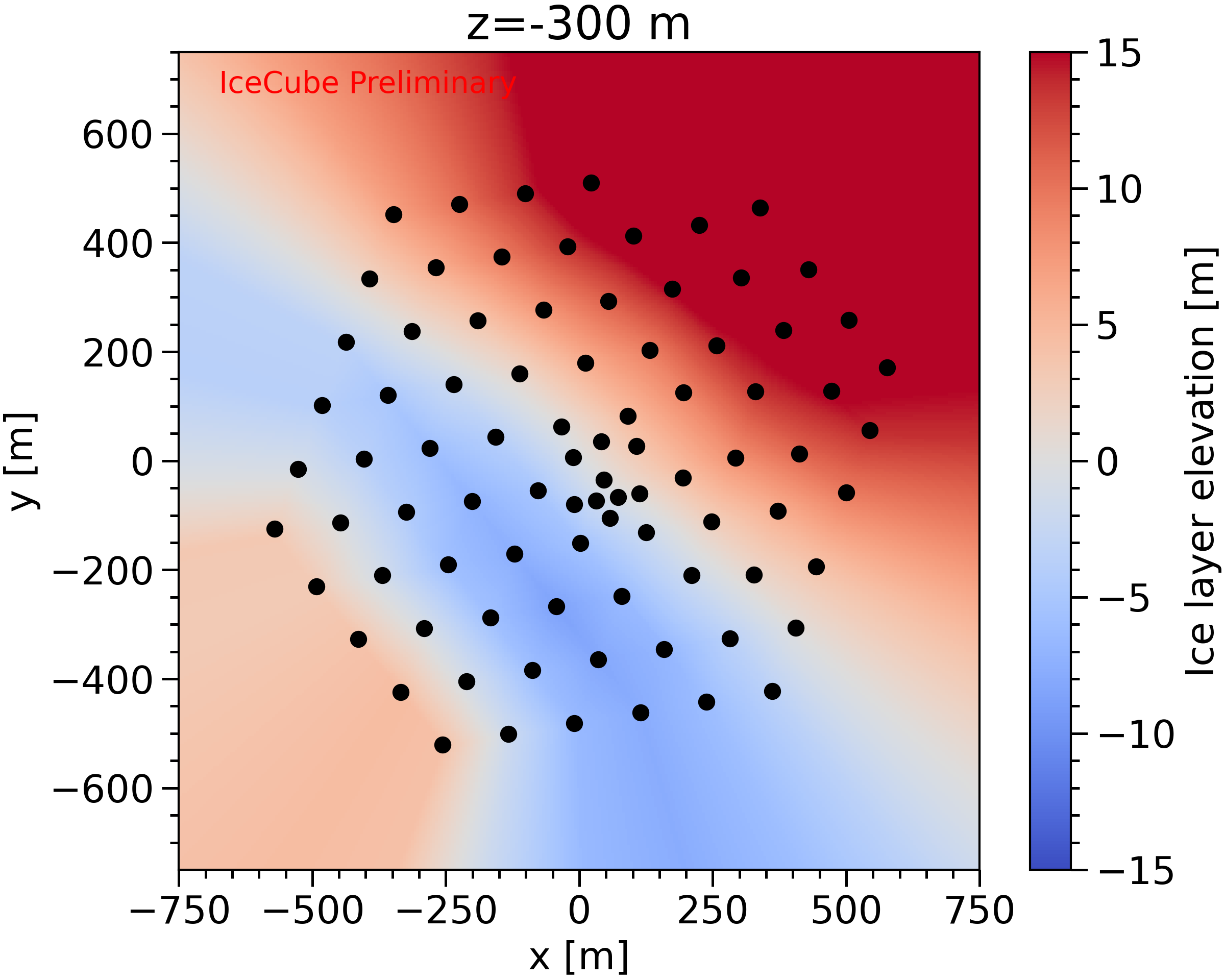}
\caption{The left (right) panel illustrates the ice layer elevation at a depth of $z=\SI{-300}{\m}$ in IceCube-centric coordinates for the simplifed (full) tilt parameterization~\cite{Chirkin:2023icrc}. Black dots indicate the $(x,y)$ positions of IceCube strings. For both panels, the color indicates the layer elevation change relative to a reference location near the origin. Note the region outside the instrumented footprint is extrapolated.}
\label{fig:tilt}
\end{figure*}

Polar ice stratigraphy broadly exhibits a strong dependence on depth, $z$, reflecting changes in Earth's climate over time. In IceCube, ice properties are defined as a function of $z$ at a fixed reference location, which is near the origin in the horizontal $xy$-plane~\cite{IceCube:2013llx,Chirkin:2023icrc}. However, the ice layers are not perfectly flat. Figure~\ref{fig:tilt} shows the ice layer elevation change across the $xy$-plane relative to the $z=\SI{-300}{\m}$ layer at the reference location. The right (left) panel shows the ice layer elevation for a full (simplified) description of the ice tilt~\cite{Chirkin:2023icrc}. This provides a mapping of the tilt correction for photon propagation Monte Carlo (MC) to look up calibrated scattering and absorption coefficients for any position throughout the detector. As a photon propagates through the ice, its position is used to look up an effective depth, $\zff(x,y,z)$, corresponding to the appropriate ice layer at the reference location. Fast reconstruction routines do not track photons individually, but can approximate the effect of tilt by using the source cascade position instead to obtain a tilt-corrected depth for the cascade, which is then used to derive expected photon yields. In addition, as the correction is linearly interpolated between fixed points in position space, we can compute the Jacobian terms $\partial \zff / \partial x_i$, where $x_i \in \{x, y, z\}$, for use in gradient-based minimizers\footnote{The linear interpolation means the tilt correction is not everywhere smooth but in practice this does not pose an issue as discontinuities in its derivatives are small and only rarely occur.}.

Factorizing the tilt correction out from the splined model itself makes it extremely simple to switch to newer tilt models for use in reconstruction. If, instead, one were to generate photon-yield expectations without this factorization, then for each update to the tilt model a massive resimulation campaign would be needed to generate the raw data for building the approximate models. In addition, the broken symmetry in $x$-$y$ would introduce two additional input dimensions, increasing the complexity of the problem. The cost is that only using the source position of the cascade is an approximation, as the path between source and receiver (DOM) can traverse several layers of ice all with varying amounts of tilt. Fortunately, the ice layer undulations are gradual and this approximation improves for shorter source-receiver distances, which is also where most of the statistics used in reconstruction is expected.

\section{Bechmarking with cascade simulation}\label{sec:benchmarks}

As mentioned in \textsection\ref{sec:intro}, cascades are primitives for reconstruction. It is, first and foremost, important to ensure that their modeling is accurate. This can be achieved by comparing reconstruction performance for a sample of MC simulated particle showers. One common metric to evaluate such performance is the median angular resolution, and it has been shown that the improvement in cascade angular resolution over an older model is about a factor of two~\cite{YUAN2023168440}. Another metric for comparison is the negatived log-likelihood per degree of freedom (reduced negative log-likelihood), $\iota$, obtained by comparing expected photon yields from the best-fit cascade to the simulated data. Here we use an effective Poisson-based likelihood, Eq.~(3.16) from~\cite{Arguelles:2019izp}, modified such that the $\sigma$ term is fixed to a relative fraction of $\mu$. Instead of representing the MC statistical uncertainty, $\sigma$ can be interpreted as an overall relative model uncertainty that broadens the shape of the likelihood space, which can be due to some combination of statistical fluctuations, fitting error and model approximations.
\begin{figure*}[hbt]
\centering
\includegraphics[width=0.6\textwidth]{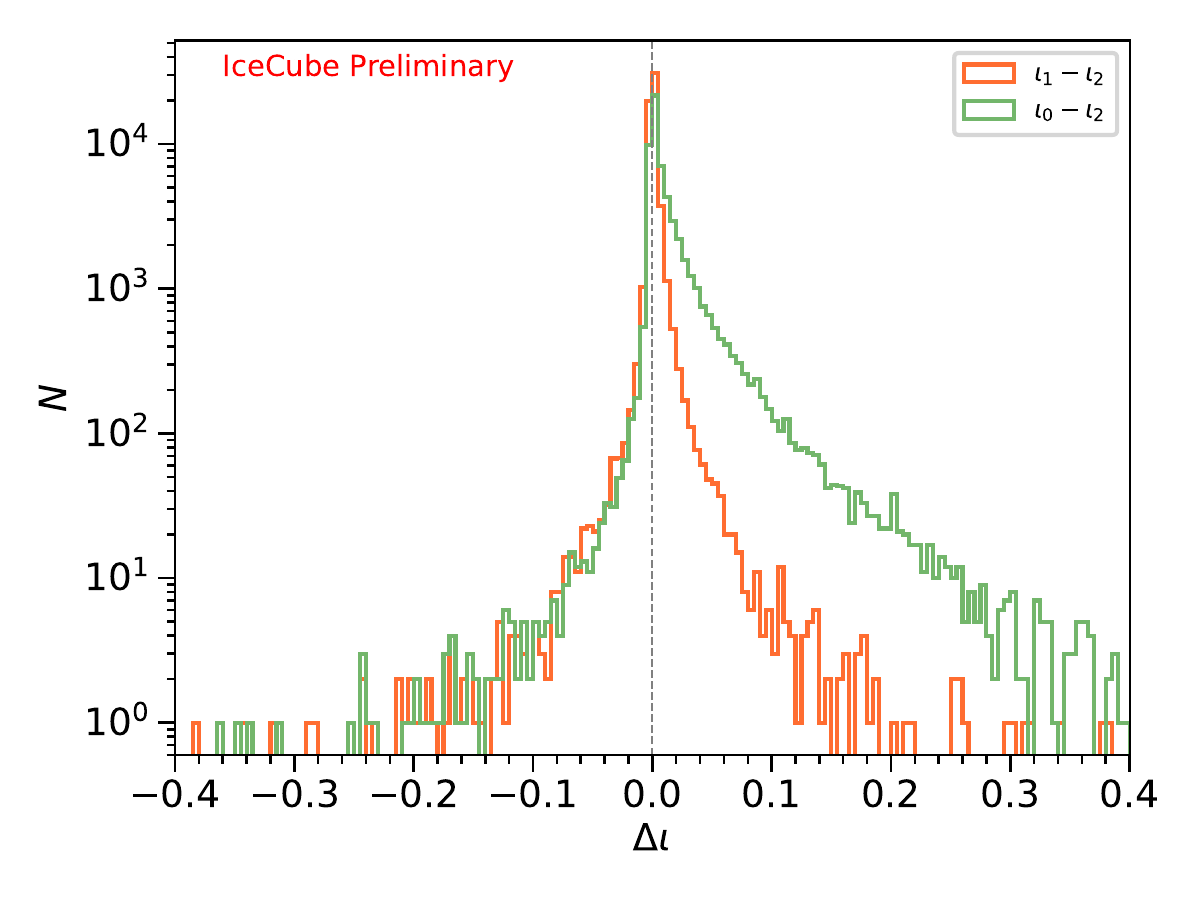}
\caption{The impact on $\iota$ due to the tilt correction, based on a simulation that includes a three dimensional description of ice layer undulations. Shown are distributions across $\Delta \iota$ for individual events, $\iota_2$ ($\iota_1$) refers to the reduced negative log-likelihood for a reconstruction that employs the full (simplified) tilt correction map, and $\iota_0$ when tilt correction is not applied in reconstruction. Lower $\iota$ values corresponds to improved agreement with the simulated data. The improved description with $\iota_2$ is evidenced by the skew towards positive values, which is more apparent for $\iota_0 - \iota_2$ (green) but still visible for $\iota_1 - \iota_2$ (orange).}
\label{fig:iota}
\end{figure*}

Figure~\ref{fig:iota} illustrates the $\iota$ improvement when using the full tilt parameterization (FTP) to obtain $\zff$ in order to reconstruct a simulated sample of showers that primarily interact within the instrumented region of IceCube. The simulation is produced with the FTP model as well~\cite{Chirkin:2023icrc}. Shown are two $\Delta \iota$ distributions, $\iota_1 -\iota_2$ and $\iota_0 - \iota_2$, where $\iota_{0,1,2}$ corresponds to the reduced negative log-likelihood for reconstructions that assume no tilt, a simplified tilt correction, and FTP correction, respectively. As more negative values indicate improved event description when compared to the simulated data, the positive skew observed for both distributions indicates the improvement of $\iota_2$ over $\iota_1$ and $\iota_0$. As expected, the improvement over $\iota_0$ is much larger than that over $\iota_1$ as the differences between the FTP model and the simplified tilt model are smaller than the values of the tilt corrections themselves (c.f. Fig.~\ref{fig:tilt}).

\section{Applications for realtime alerts}\label{sec:realtime}

\begin{figure*}[hbt]
\centering
\includegraphics[width=0.49\textwidth]{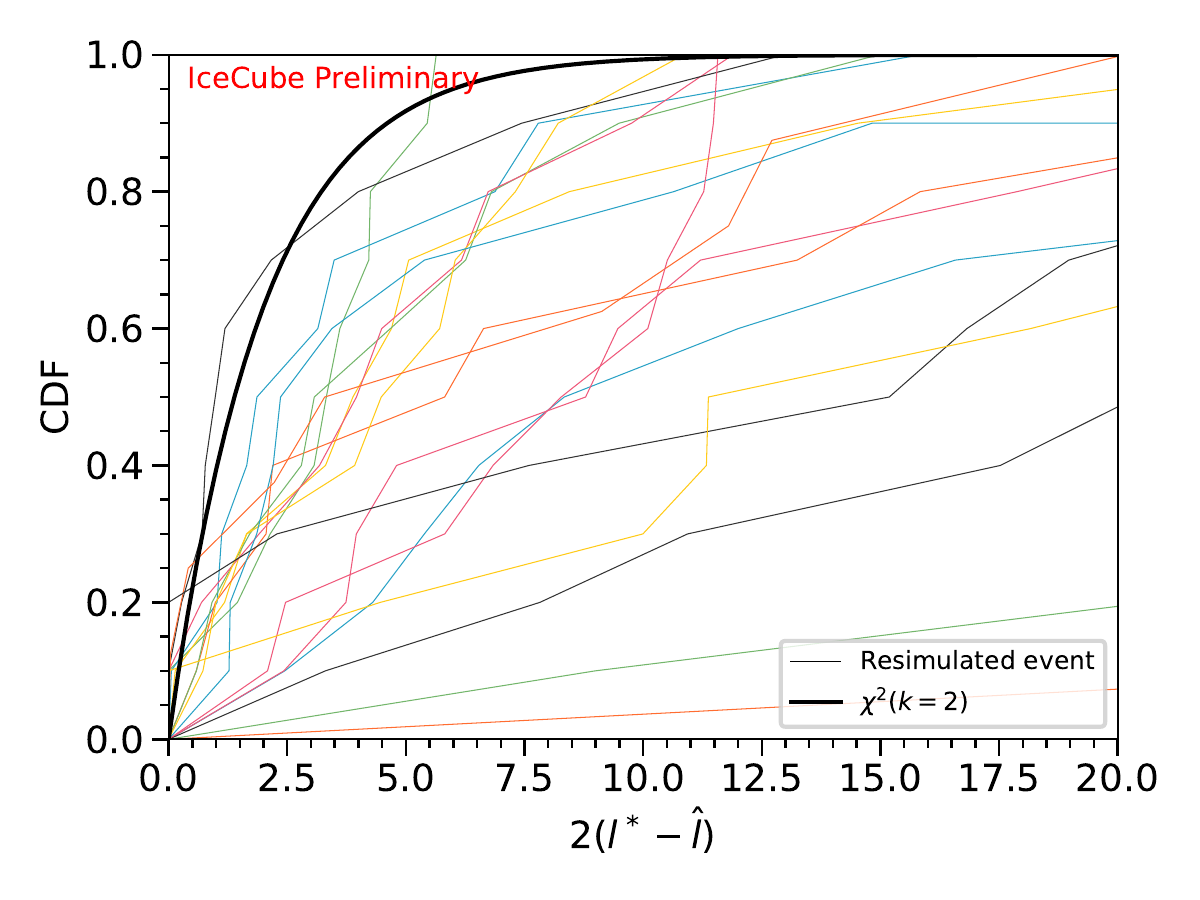}
\includegraphics[width=0.49\textwidth]{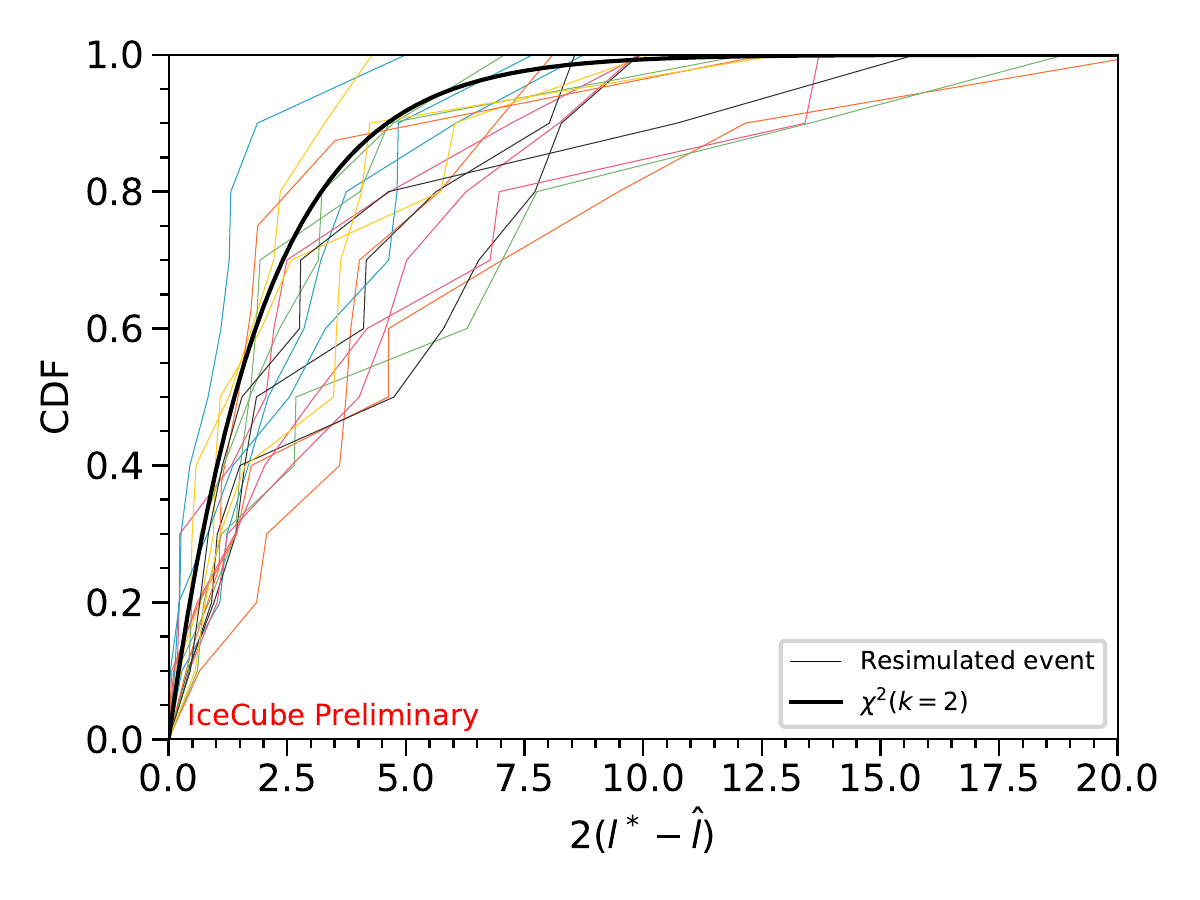}
\caption{The left (right) panel shows the cumulative distribution function (CDF) obtained using the existing (updated) reconstruction algorithm for a small sample of simulated realtime tracks. The black line indicates the CDF of a chi-square distribution with two degrees of freedom. Each colored line is obtained for the same underlying event, simulated multiple times with varying ice properties and statistical fluctuations, and shows the CDF of $2(l^*-\hat{l})$, where $l^*$ and $\hat{l}$ are the negative log-likelihoods at the true and best-fit directions, respectively. The same set of events are plotted in both panels, and an improved convergence to $\chi^2(k=2)$ is observed in the right panel when the updates detailed in \textsection\ref{sec:realtime} are incorporated into the reconstruction.}
\label{fig:chi2}
\end{figure*}

Building on the improvements observed in cascade modeling discussed here and in~\cite{YUAN2023168440}, the next step was to apply it to existing track reconstruction of high-energy muons. Such muons lose energy via stochastic processes~\cite{Koehne:2013gpa}, which can in turn be reconstructed as a series of aligned and equally-spaced cascades~\cite{Aartsen:2013vja}. More accurate cascade modeling is therefore directly applicable for track reconstruction. In particular, realtime alerts in IceCube are reconstructed under this assumption with an iterative scan of the likelihood performed over the sky. One serious drawback of the approach currently taken in realtime scans is that the likelihood space does not conform to expectations from Wilks' theorem. Resimulation studies have shown that level contours based on the log-likelihood space exhibit large event-to-event variance (c.f. left panel Fig.~\ref{fig:chi2}), making it challenging to provide realistic per-event confidence regions in direction. A main goal of the work described here is to implement updates in modeling and reconstruction in order to converge towards the Wilksian limit.

Several complications can arise in the reconstruction of high-energy muons. The cascade model used in reconstruction assumes a point-like shower while in reality particle showers have longitudinal extension on the order of a few meters depending on the energy~\cite{Radel:2012ij}. Additionally, the photon flux approaches a singularity as the source-receiver distance approaches zero, making it challenging to describe with spline fits. Thus, the near-field regime where the source-receiver distance is small is inaccurately modeled. To work around this, typically the highest-charge DOMs are excluded from reconstruction. In the case of tracks traveling through the detector, there is a higher probability of the track passing near DOMs along its path. The threshold for the upper bound on kept DOM charges was lowered to exclude additional DOMs, which are expected to be near muon stochastic losses.

Another factor that can affect track reconstruction for realtime is the minimizer step direction in position space. As a brute-force scan is performed over a set of directions, for each fixed direction,  $(\zen, \azi)$, the likelihood needs to be minimized over a vertex position that defines the track location. For a fixed direction and a throughgoing track hypothesis, the likelihood space along the longitudinal axis aligned with the track is essentially flat, since a shift forward or backwards along an infinite line results in the same line. To better exploit this symmetry, steps in position space can be projected onto a plane orthogonal to the track direction, and the minimizer is setup to prefer larger transverse steps in the plane than along the longitudinal axis. This change resulted in improved likelihood minimization with much fewer vertex seeds needed to escape local minima.

In addition to a reduction in the number of vertex seeds, a computational speed-up was achieved by updating the treatment of unhit DOMs. By default, all unhit DOMs are included in reconstruction as a lack of detection still contains information. However, this forces a loop over more than 5000 DOMs at each iteration of the minimization. On the other hand, excluding all unhit DOMs would discard information while possibly introducing local minima. Instead, a natural compromise is to incorpoate a sparser array of unhit DOMs, for example keeping only one out of every five. Such an approach yields a speed up in track reconstruction while keeping information contained in the kept unhit DOMs. In conjunction with some tuning of the binning in time, local minima can be avoided.

Figure~\ref{fig:chi2} shows a comparison of the cumulative distribution function (CDF) of differences in the negative log-likelihood, $l$, for the existing (left panel) and updated (right panel) reconstruction routines. Each CDF (colored line) was obtained by repeated simulation of the same underlying track event with varying ice properties and statistical fluctuations. Then for each resimulation, a full-sky likelihood scan was performed and the difference between the negative log-likelihood at the true direction, $l^*$, and at the best-fit direction, $\hat{l}$, is obtained. The CDF of $2*(l^*-\hat{l})$ is then constructed and compared to that of a $\chi^2$ distribution with two degrees of freedom (black), to which it should ideally converge. Due to the computational time-complexity of full-sky scans only a small sample of O(10) resimulations per event is processed. With additional resimulations reconstructed the per-event CDFs is expected to smoothen out. The CDFs obtained with the reconstruction routine that incorporates the updates described here (right panel) is a clear improvement in agreement to $\chi^2(k=2)$ over the extant reconstruction (left panel), though the convergence is imperfect, indicating additional room for improvement.

\section{Outlook and conclusion}\label{sec:outro}

As glacial ice modeling evolves and new calibration of the ice becomes available, it is necessary to update models used in reconstruction as well. By factorizing the tilt correction out from splined fits, tilt corrections can be applied to reconstruction in a similar fashion as in simulation itself. While this is an approximation, the studies shown in \textsection\ref{sec:benchmarks} show that it is a reasonable one to make. By applying the FTP tilt correction in reconstruction, an improved agreement with fully simulated showers is obtained.

Some limitations of the cascade modeling related to small source-receiver distances have already been discussed in \textsection\ref{sec:realtime}. One remaining limitation lies in the accurate modeling of arrival-time probability density functions (PDF)  for anisotropic ice. The effective distance correction is constructed by comparing time-integrated photon yields for an anisotropic ice model to one without ice anisotropy~\cite{YUAN2023168440}. It therefore sets a correction to the overall normalization, but does not capture the behavior of time PDFs for anisotropic ice. In order to obtain a full anisotropic description of the time PDF requires either adding new dimensions to the model, namely $\azi$, or perhaps a reparameterization to a different space. The former is currently not computationally tractable. Machine learning approaches have also shown promise~\cite{Glusenkamp:2023icrc,IceCube:2021umt} but is beyond the scope of this discussion.

\bibliographystyle{ICRC}
\bibliography{main}

%

\clearpage

\section*{Full Author List: IceCube Collaboration}

\scriptsize
\noindent
R. Abbasi$^{17}$,
M. Ackermann$^{63}$,
J. Adams$^{18}$,
S. K. Agarwalla$^{40,\: 64}$,
J. A. Aguilar$^{12}$,
M. Ahlers$^{22}$,
J.M. Alameddine$^{23}$,
N. M. Amin$^{44}$,
K. Andeen$^{42}$,
G. Anton$^{26}$
% ...
\\
\\
% ...
$^{12}$ Universit{\'e} Libre de Bruxelles, Science Faculty CP230, B-1050 Brussels, Belgium \\
% ...
$^{17}$ Department of Physics, Loyola University Chicago, Chicago, IL 60660, USA \\
$^{18}$ Dept. of Physics and Astronomy, University of Canterbury, Private Bag 4800, Christchurch, New Zealand \\
% ...
$^{22}$ Niels Bohr Institute, University of Copenhagen, DK-2100 Copenhagen, Denmark \\
$^{23}$ Dept. of Physics, TU Dortmund University, D-44221 Dortmund, Germany \\
% ...
$^{26}$ Erlangen Centre for Astroparticle Physics, Friedrich-Alexander-Universit{\"a}t Erlangen-N{\"u}rnberg, D-91058 Erlangen, Germany \\
% ...
$^{40}$ Dept. of Physics and Wisconsin IceCube Particle Astrophysics Center, University of Wisconsin{\textendash}Madison, Madison, WI 53706, USA \\
% ...
$^{42}$ Department of Physics, Marquette University, Milwaukee, WI, 53201, USA \\
% ...
$^{44}$ Bartol Research Institute and Dept. of Physics and Astronomy, University of Delaware, Newark, DE 19716, USA \\
% ...
$^{63}$ Deutsches Elektronen-Synchrotron DESY, Platanenallee 6, 15738 Zeuthen, Germany  \\
$^{64}$ Institute of Physics, Sachivalaya Marg, Sainik School Post, Bhubaneswar 751005, India \\
% ...

\subsection*{Acknowledgements}

\noindent
The authors gratefully acknowledge the support from the following agencies and institutions:
USA {\textendash} U.S. National Science Foundation-Office of Polar Programs,
U.S. National Science Foundation-Physics Division,
U.S. National Science Foundation-EPSCoR,
Wisconsin Alumni Research Foundation,
Center for High Throughput Computing (CHTC) at the University of Wisconsin{\textendash}Madison,
Open Science Grid (OSG),
Advanced Cyberinfrastructure Coordination Ecosystem: Services {\&} Support (ACCESS),
Frontera computing project at the Texas Advanced Computing Center,
U.S. Department of Energy-National Energy Research Scientific Computing Center,
Particle astrophysics research computing center at the University of Maryland,
Institute for Cyber-Enabled Research at Michigan State University,
and Astroparticle physics computational facility at Marquette University;
Belgium {\textendash} Funds for Scientific Research (FRS-FNRS and FWO),
FWO Odysseus and Big Science programmes,
and Belgian Federal Science Policy Office (Belspo);
Germany {\textendash} Bundesministerium f{\"u}r Bildung und Forschung (BMBF),
Deutsche Forschungsgemeinschaft (DFG),
Helmholtz Alliance for Astroparticle Physics (HAP),
Initiative and Networking Fund of the Helmholtz Association,
Deutsches Elektronen Synchrotron (DESY),
and High Performance Computing cluster of the RWTH Aachen;
Sweden {\textendash} Swedish Research Council,
Swedish Polar Research Secretariat,
Swedish National Infrastructure for Computing (SNIC),
and Knut and Alice Wallenberg Foundation;
European Union {\textendash} EGI Advanced Computing for research;
Australia {\textendash} Australian Research Council;
Canada {\textendash} Natural Sciences and Engineering Research Council of Canada,
Calcul Qu{\'e}bec, Compute Ontario, Canada Foundation for Innovation, WestGrid, and Compute Canada;
Denmark {\textendash} Villum Fonden, Carlsberg Foundation, and European Commission;
New Zealand {\textendash} Marsden Fund;
Japan {\textendash} Japan Society for Promotion of Science (JSPS)
and Institute for Global Prominent Research (IGPR) of Chiba University;
Korea {\textendash} National Research Foundation of Korea (NRF);
Switzerland {\textendash} Swiss National Science Foundation (SNSF);
United Kingdom {\textendash} Department of Physics, University of Oxford.

\end{document}